# Solid-State Li-Ion Batteries Operating at Room Temperature Using New Borohydride Argyrodite Electrolytes

Anh Ha Dao [1], Pedro López-Aranguren [1,2], Junxian Zhang [1], Fermín Cuevas [1,*] and Michel Latroche [1]

[1] department/school/faculty/campus, Univ. Paris Est Creteil, CNRS, ICMPE, UMR7182, 7182, 2 rue Henri Dunant, F-94320 Thiais, France; daohaanh1988@gmail.com (A.H.D.); plopez@cicenergigune.com (P.L.-A.); junxian@icmpe.cnrs.fr (J.Z.); latroche@icmpe.cnrs.fr (M.L.)

[2] Center for Cooperative Research on Alternative Energies (CIC energiGUNE), Basque Research and Technology Alliance (BRTA), Parque Tecnológico de Álava, Albert Einstein, 48, 01510 Vitoria-Gasteiz, Spain

* Correspondence: cuevas@icmpe.cnrs.fr; Tel.: +33-149-781-225

**Abstract:** Using a new class of $(BH_4)^-$ substituted argyrodite $Li_6PS_5Z_{0.83}(BH_4)_{0.17}$, (Z = Cl, I) solid electrolyte, Li-metal solid-state batteries operating at room temperature have been developed. The cells were made by combining the modified argyrodite with an In-Li anode and two types of cathode: an oxide, $Li_xMO_2$ ($M$ = ⅓ Ni, ⅓ Mn, ⅓ Co; so called NMC) and a titanium disulfide, $TiS_2$. The performance of the cells was evaluated through galvanostatic cycling and Alternating Current AC electrochemical impedance measurements. Reversible capacities were observed for both cathodes for at least tens of cycles. However, the high-voltage oxide cathode cell shows lower reversible capacity and larger fading upon cycling than the sulfide one. The AC impedance measurements revealed an increasing interfacial resistance at the cathode side for the oxide cathode inducing the capacity fading. This resistance was attributed to the intrinsic poor conductivity of NMC and interfacial reactions between the oxide material and the argyrodite electrolyte. On the contrary, the low interfacial resistance of the $TiS_2$ cell during cycling evidences a better chemical compatibility between this active material and substituted argyrodites, allowing full cycling of the cathode material, 240 mAhg$^{-1}$, for at least 35 cycles with a coulombic efficiency above 97%.

## 1. Introduction

Besides other storage means, to date, batteries are considered among the best candidates to store energy. In particular, nowadays, the Li-ion technology offers the best compromise between reliability, cost, and energy density. However, developments are still needed for this latter technology to fulfill the energy density targeted by the market of electric vehicles (EV). In this regard, the solid-state battery (SSB) is considered the most promising candidate providing high energy and power with improved safety. Indeed, SSB are expected to mitigate some hazards related to the use of current liquid organic electrolytes (flammability, dendrite formation with Li, Solid Electrolyte Interface (SEI) growth) and to allow for thinner electrolyte layers, allowing higher power and energy densities. The enhancement on the performance of SSBs ties in closely with the development of solid electrolytes (SE), providing a high ionic conductivity, a sufficient electrochemical stability, and chemical compatibility with the active materials [1–4].

Among the range of inorganic SEs, sulfides are excellent candidates due to their high ionic conductivity (as high as $10^{-2}$ S cm$^{-1}$) at room temperature (RT) and high lithium transference number (~1) [5]. Besides, their remarkably low Young modulus (~20 GPa) facilitates the assembly of the cell components at low mechanical pressures and ensures a low electrolyte–electrode interfacial resistance [6,7]. One of the most relevant features of SSBs is the high energy density, reachable with high-voltage active materials and enabled by the solid electrolyte [8,9]. The NMC is a high-voltage active material that has been used in SSBs with different sulfide-based solid electrolytes such as $80Li_2S \cdot 19P_2S_5 \cdot 1P_2O_5$ [10], $Li_3PS_4$ (LPS) [11,12], $Li_{10}GeP_2S_{12}$ (LGPS) [13] and argyrodites with the formula $Li_6PS_5Z$ (Z = Cl, Br, I) [14,15]. The electrochemical performance of these batteries initially displayed theoretical capacities fading to ca. 50 mAhg$^{-1}$ after a few hundreds of cycles. The poor performance of these devices is ascribed to the de-oxidation of the sulfide by the active material and loss of contact between the particles after cycling [16]. Coating of NMC with an oxide as protective layer is a common strategy to decrease the degradation of the sulfide and improve the life cycle of the cells [11]. A superior performance has been reported with other active materials such as $TiS_2$. For instance, Kanno et al. showed a reversible capacity of 160 mAhg$^{-1}$ at 1$C$ with an LGPS cell [17]. The combination of LGPS and LPS as solid electrolyte is also highlighted by a cell attaining 25% of the theoretical capacity at 20 C [18]. In spite of this, the composition of the sulfide electrolyte for solid state devices has to be carefully chosen in order to avoid degradation occurring at the high-voltage active material/electrolyte interface [19].

Recently, borohydride-based solid electrolytes have been reported showing high ionic conductivity near room temperature [20,21]. This novel family of materials opens promising routes to develop safe and efficient solid electrolyte. The next step consists in building full batteries associating working electrodes with these new electrolytes [22].

In a previous work, we reported a new type of argyrodite in which the halide $Z^-$ ions are partially substituted by $(BH_4)^-$ units. The SE is prepared by mechanical milling of argyrodite precursors ($Li_2S$, $P_2S_5$, and LiZ) and $LiBH_4$, with a molar ration $LiBH_4$/LiZ of 1/6 [23]. The molecular structure of these compounds is provided in [23]. The high conductivity (between 4.1 and 7.6×10$^{-4}$ Scm$^{-1}$ at RT) and wide electrochemical stability up to 5 V vs. Li/Li$^+$ of these $(BH_4)^-$ substituted argyrodites turn them out into great candidates for their application in high-voltage SSBs performing at RT. The present work describes an electrochemical study conducted at RT on SSBs using two different argyrodite-type electrolytes in which the halogen (Cl or I) is partially substituted by $BH_4$. The electrochemical cells are assembled using these electrolytes, an indium–lithium based anode and two different cathodes, either NMC or $TiS_2$, which operate at high and medium voltage, respectively.

## 2. Materials and Methods

*2.1. Solid-State Electrolytes*

Reagent-grade $Li_2S$ (Sigma Aldrich, Saint Louis, MO, USA, 99.98%), $P_2S_5$ (Sigma Aldrich, Saint Louis, MO, USA, 98%), $LiBH_4$ (Rockwood Lithium, Geesthacht, Germany, 97.8%), LiCl, and LiI (Sigma Aldrich, Saint Louis, MO, USA, 99.99%) were used as precursors. Powders were mixed in stoichiometric ratio to form the nominal compositions $Li_6PS_5Cl_{0.83}(BH_4)_{0.17}$ (Cl-SE) and $Li_6PS_5I_{0.83}(BH_4)_{0.17}$ (I-SE). To prepare the electrolytes, 1 g of mixture of the reactants was milled under argon atmosphere in a planetary ball-milling equipment (Pulverisette 7, Fritsch, FRITSCH GmbH Idar-Oberstein, Germany ) at 600 rpm for 20 h using 25 stainless balls ($\rho$ = 7.8 g cm$^{-3}$) of 7 mm in diameter. More details about SE preparation can be found elsewhere [23].

*2.2. Cathode Materials*

For the high-voltage cell, commercial Li$M$O$_2$; $M$ = ⅓ Ni, ⅓ Mn, ⅓ Co (NMC, Sigma-Aldrich, Saint Louis, MO, USA, 98%) was used as active material. The practical capacity for NMC is 175 mAhg$^{-1}$ (i.e., 0.6 Li) due to the non-uniform conductivity of Li$_x$$M$O$_2$ at different states of charge [24]. Owing to the low intrinsic electrical conductivity of NMC (10$^{-7}$ S cm$^{-1}$ at discharged state) [25], conductive carbon black C65 was added to the cathode composite. The electrode was prepared by hand mixing of NMC, C65, and the SE in an agate mortar with ratio fixed to 25:5:70 wt.%, respectively. For the preparation of the middle-voltage electrode, titanium disulfide (TiS$_2$, Sigma Aldrich, Saint Louis, MO, USA, 99.9%) powder was used as active material (240 mAhg$^{-1}$). Due to its intrinsic high conductivity (3.3×10$^3$–4.7×10$^2$ S cm$^{-1}$ [26,27], no carbon conductive agent was needed. The electrode was prepared by mortar hand mixing of TiS$_2$ with the solid electrolyte (SE) in the TiS$_2$:SE ratio 30:70 wt.%. For both active materials, the electrolyte volume fraction was set around 0.8, ensuring that the ion transport pathways would not be a limiting factor in these composite electrodes [28].

*2.3. Cell Assembly*

Solid-state cells were assembled by pelletizing together 3–5 mg of the cathode mixture and 80 mg of the solid electrolyte (either Cl-SE or I-SE) in a 10 mm diameter die applying 2 tons. A bilayer of Li and In foil (250 μm and 100 μm thickness) was placed with the In-side facing the solid electrolyte. The Li-In|SE|cathode materials were assembled in a home-made sealed Teflon Swagelok cell and conditioned overnight at 75 °C to form the In-Li alloy (35 Li:65 In at.%) [29,30] and to minimize the interfacial resistance of the cells.

*2.4. Electrochemical Measurements*

The electrochemical performance of the cells was evaluated by galvanostatic cycling. The rate capability of the NMC and TiS$_2$ cells was evaluated on a multichannel battery tester (MPG-2, Biologics, Seyssinet-Pariset, FRANCE ) in the 2.0–4.0 V and 0.8–2.2 V range, respectively, at the following *C*-rates: 0.02 *C*, 0.2 *C*, 0.1 *C* back to 0.02 *C* for NMC; 0.01*C*, 0.5*C*, 1*C* back to 0.01 *C* for TiS$_2$. The NMC cell was cycled at 25 °C and 40 °C to study the impact of the temperature upon cycling. The In-Li anode has been used to decrease the reduction potential by 0.6 V vs. Li/Li$^+$ thus limiting the reactivity with TiS$_2$ [29,30]. The capacity is reported with respect to the mass of the positive active material. AC Electrochemical Impedance Spectroscopy (EIS) measurements were acquired at 25 °C using an Autolab PGSTAT30 potentiostat ( Metrohm Villebon Courtaboeuf, FRANCE ). The input voltage perturbation and the frequency range were set to 10 mV and 1MHz–1Hz, respectively. The data were processed with ZView® software (V3.4, Scribner, Southern Pines, NC USA).

## 3. Results

*3.1. High-Voltage NMC│SE│In-Li Cell*

Figure 1a shows the discharge rate capabilities for the NMC cells, including Cl-SE and I-SE as solid electrolytes. At 0.02 *C* and 25 °C, the cell including I-SE delivered 75 mAhg$^{-1}$. At faster *C*-rates the capacity decreased down to 25 and 10 mAhg$^{-1}$. A discharge capacity of 60 mAhg$^{-1}$ was recovered at 0.02 *C*, evidencing a rapid capacity fading of the cell. A similar behavior was observed for the cell including Cl-SE, although the initial discharge capacity for this cell reached 100 mAhg$^{-1}$ and a capacity 50% higher than the I-SE cell at faster *C*-rates. The performance of the cells evidenced kinetics issues arising from the difference on the ionic conductivity of the electrolytes: the I-SE ionic conductivity at 25 °C was almost twice that of Cl-SE (7.6 × 10$^{-4}$ and 4.1 × 10$^{-4}$ Scm$^{-1}$, respectively) [23]. The NMC|Cl-SE|In-Li cell was cycled at 40 °C to elucidate the as-mentioned kinetic limitations. Figure 1a shows how at 40 °C, the cell almost reached the theoretical capacity, with a capacity fading of ~7%, over 10 cycles at 0.02 *C*. This value compares well with the literature [24]. In addition, the voltage profile comparison at 25 and 40 °C (Figure 1b) displayed a lower polarization of the cell cycled at 40 °C, as a result of the higher conductivity of Cl-SE at 40 °C. The recovery of the cell capacity at low *C* rates agreed with ionic diffusion limitations rather than considering chemical side reactions. The coulombic efficiency (CE) of the cells (not shown here) varied between 75 and 90% depending on the *C*-rate. Such low CE has been ascribed to a continuous degradation of the cell from irreversible side reactions between NMC and the TiS$_2$ during the oxidation (charge) step (Figure 1b) [16].

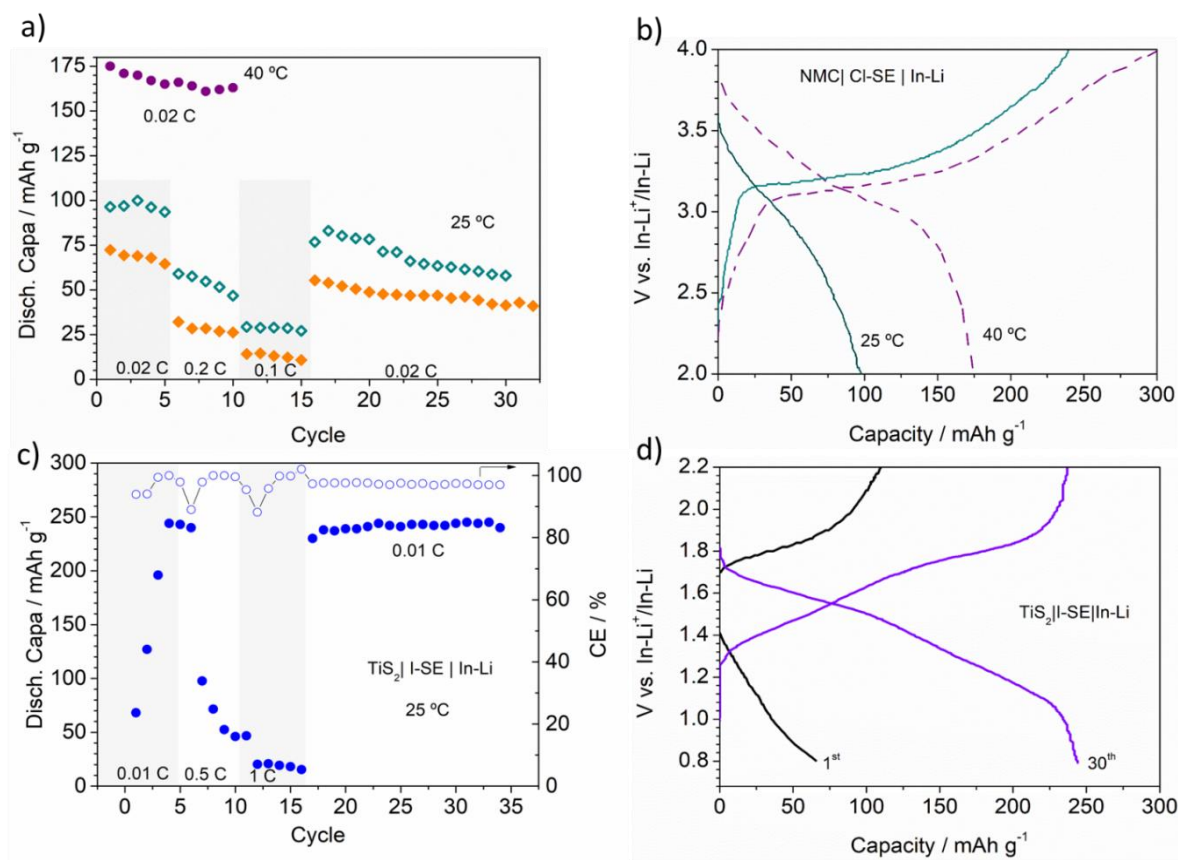

**Figure 1.** (**a**) Rate discharge capability of the NMC|Cl-SE|In-Li cell at 25 °C (green) and 40 °C (purple), and of the NMC|I-SE|In-Li cell at 25 °C (orange). (**b**) Charge/discharge profiles of the NMC|Cl-SE|In-Li cell obtained at the initial state (0.02 C) at 25 °C (solid line) and 40 °C (dotted line). (**c**) Rate discharge capability and coulombic efficiency (CE) of the TiS$_2$|I-SE|In-Li cell at 25 °C. (**d**) Initial and 30th charge/discharge profile of the TiS$_2$|I-SE|In-Li cell at 25 °C.

The EIS spectra of the solid-state cell NMC|Cl-SE|In-Li measured at 25 °C before and after 20 cycles at the charged state are shown in Figure 2a. Before cycling the cell, the EIS showed a clear semicircle obtained at high frequency (1 MHz–1 kHz), followed by a partial semicircle at middle frequency (100 Hz–1 Hz). The spectra have been fitted to the equivalent circuit proposed in Figure 3. The intersection of the high frequency semicircles with the real axis, were attributed to the resistance R1 of the electrolyte and cell internal connections, 170 and 150 Ω before and after 20 cycles, respectively. The semicircle at high frequency was assigned to the charge transfer resistance at the interface between the In-Li anode and the solid electrolyte, equivalent to the parallel contribution of constant phase elements (CPE1) and diffusion-controlled Warburg impedance Zw1. The impedance Zw1 increased from 1040 Ω before cycling to 1858 Ω after 20 cycles. The partial semicircle at middle frequency was assigned to resistivity effects at the SE/cathode interface as well as within the cathode and corresponded to the parallel circuit of constant phase elements (CPE2) and diffusion controlled Warburg impedance Zw2 in Figure 3. For the as-mounted cell, Zw2 was 2000 Ω and it increased to 47 kΩ after 20 cycles.

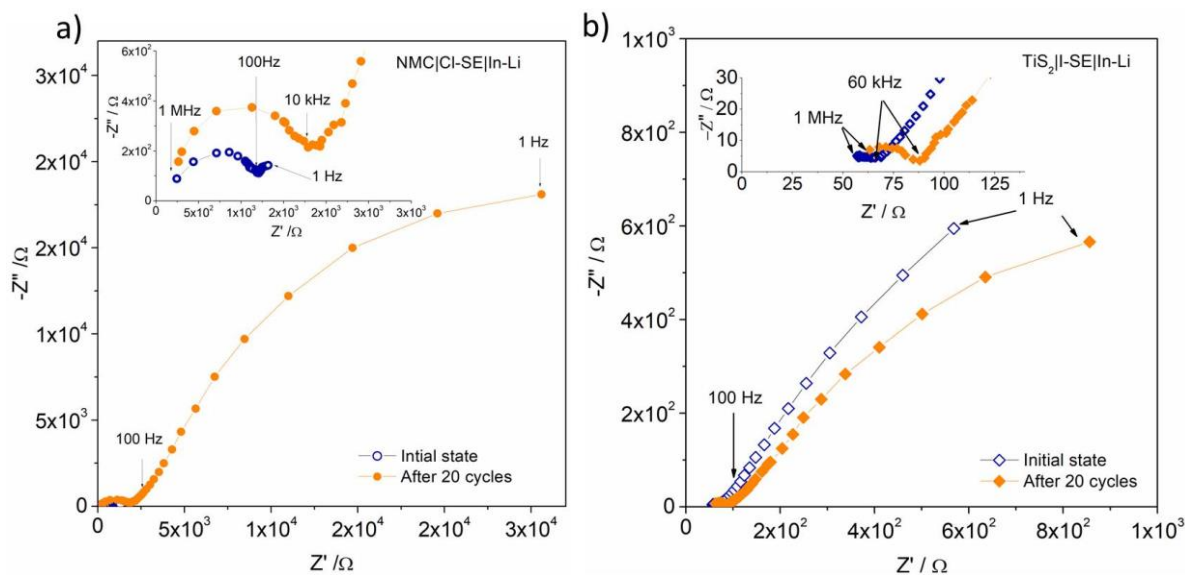

**Figure 2.** (**a**) Impedance profiles of the NMC|Cl-SE|In-Li cell obtained before and after cycling at 25 °C, (**b**) Impedance profiles of the TiS$_2$|I-SE|In-Li cell obtained before and after cycling at 25 °C.

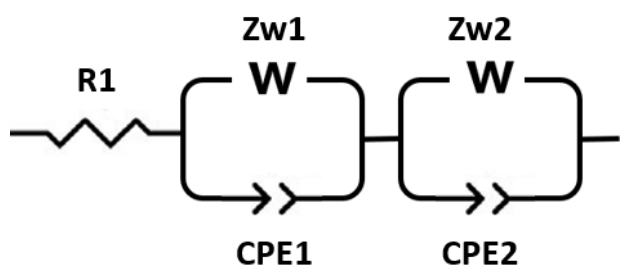

**Figure 3.** Equivalent circuit proposed for the NMC and TiS$_2$ solid state full cells.

Similar spectra (Figure S1) were obtained for the cell NMC|I-SE|In-Li and can be modeled with the same equivalent circuit (Figure 3). The R1 was around 120 Ω before and 340 Ω after 20 cycles. The first semicircle at high frequency (CPE1, Zw1) gave Zw1 = 1270 Ω and 1780 Ω before and after 20 cycles, respectively. The second circle (CPE2, Zw2) at middle frequency gave Zw2 = 56 kΩ and 63 kΩ before and after 20 cycles. These values were comparable to the Cl-SE cell. From R1 values, the ionic conductivity of Cl-SE and I-SE electrolytes were determined as 4.1 and 5.9×10$^{-4}$ Scm$^{-1}$ RT, respectively, comparable to previously-reported work [23].

*3.2. TiS$_2$│I-SE│In-Li Cell*

Figure 1c shows the rate capability and CE for the TiS$_2$|I-SE|In-Li cell cycled at 25 °C. During the first three cycles the cell showed a poor performance with unstable discharge capacity and CE. The voltage profile of the first cycle at 0.01 *C* shown in Figure 1d suggests the formation of decomposition compounds during oxidation and reduction process. After these initial formation cycles, the cell successfully delivered the full capacity (240 mAhg$^{-1}$) with a 97% CE. The voltage profile of the 4th cycle shown in Figure 1d reveals stable sloppy potential curves characteristics from TiS$_2$. At increasing *C*-rates, the capacity of the cell decreased down to 40 and 20 mAhg$^{-1}$ at 0.5 *C* and 0.1 *C*, respectively, evidencing kinetic limitations likely arising from the insufficient conductivity of the SE. Back to 0.01 *C*, the cell recovered, showing stable cycles and delivering the theoretical capacity for 35 cycles.

The EIS of the TiS$_2$|I-SE|In-Li cell was recorded at 25 °C before and after 20 cycles (Figure 2b). The spectra were fitted to the same equivalent circuit as the NMC cells (Figure 3). The intersection with the real axis at 1 MHz was around 50 Ω in both cases and assigned to the solid electrolyte

resistance R1. The same equivalent circuit as NMC was used and the Zw1 values were 170 and 118 Ω for the initial state and after 20 cycles, respectively. At low frequency, the partial semicircle was assigned to the cathode. The Zw2 reached 3348 Ω before cycling and decreased to 1394 Ω after 20 cycles. All results are summarized in Table 1 below.

**Table 1.** Electrochemical properties obtained at 25 °C for the different cells studied in this work (σ: ionic conductivity of the solid electrolyte; CE: coulombic efficiency; R1, ZW1 and ZW2 are defined in Figure 3 and given at initial state (ini.) and after twenty cycles (20 c.)).

| Cell Type | Voltage | Capacity | Capacity | σ | CE | R1 (ini./20c.) | Zw1 (ini./20 c.) | Zw2 (ini./20 c.) |
|---|---|---|---|---|---|---|---|---|
| | V | mAhg$^{-1}$ | mAhg$^{-1}$ | x10$^4$Scm$^{-1}$ | % | Ω | Ω | kΩ |
| NMC│Cl-SE│In-Li | 3.2–3.8 | 100 @ C/50 | 20 @ C/10 | 4.1 | ~90 | 170/150 | 1040/1858 | 2/47 |
| NMC│I-SE│In-Li | 3.2–3.8 | 75 @ C/50 | 10 @ C/10 | 7.6 | ~90 | 120/340 | 1270/1780 | 56/63 |
| TiS$_2$│I-SE│In-Li | 1.3–1.9 | 240 @ C/100 | 20 @ 1C | 7.6 | 97 | 50/50 | 170/118 | 3.35/1.39 |

## 4. Discussion

The SSBs using substituted argyrodites either with NMC or TiS$_2$ as active cathode materials have been cycled for up to thirty-five cycles. Several differences at the performance level of the cells were found as a function of the cathode material. Regarding potential, the measured voltages were in line with the used active materials, in the range of 3.2–3.8 V for NMC and 1.3–1.9 V for TiS$_2$ (vs. In-Li$^+$/In-Li). Both batteries presented plateau-like behavior though sloping effects were observed. For NMC, this can be related to the interface between the oxide cathode and the sulfide electrolyte, where a space-charge layer can form as result of their different chemical potentials. This layer creates a large interfacial resistance, leading to a potential slope prior to the delithiation plateau of the oxide electrode in the charge profile as reported in [30,31]. It has also been shown that at the contact surface between lithium oxide cathode, Li*MO*$_z$, and argyrodite SE, the solid electrolyte decomposes after 20 to 25 cycles into sulfides Li$_2$S$_n$, sulfur, and P$_2$S$_5$. This yields to severe fading from the first tens of cycles for NMC-argyrodite solid state cells [19]. Mutual diffusion of Co and P atoms in this interface layer is also reported as another explanation for the high resistance of oxide-sulfide cells [32]. Indeed, transition metal atoms were found in the sulfide solid electrolyte even beyond the interface [33–35]. In contrast, possible intermixing effects are expected to be less harmful for TiS$_2$/argyrodite-type interfaces due to their common sulfur chemistry. The excellent electrochemical performance found for the TiS$_2$ cell evidenced a better compatibility between the active material and the sulfide electrolyte in comparison to NMC. This is in concurrence with the cycling reported by Sakuda et al. in TiS$_2$│argyrodite│Li cells [36] and Unemoto et al. in TiS$_2$│90LiBH$_4$:10P$_2$S$_5$│InLi [37].

Using the same anode and electrolyte, it is worth comparing the behavior of the two studied active materials in the light of their different chemistry. TiS$_2$ behaved much better than NMC in SSBs using the borohydride-substituted argyrodites as electrolyte. Looking to the EIS measurements, R1 and Zw1 showed comparable values for NMC (with both Cl- and I-SE) and TiS$_2$ (with I-SE) that agreed well with the assignment of R1 to the electrolyte and Zw1 to the charge transfer resistance at the anode. This was not the case for Zw2, and such differences were indeed caused by the cathode counterpart for which the resistive values Zw2 were much higher for NMC (up to 62 kΩ) than for TiS$_2$ (around 3 kΩ). Such a difference can be attributed to two phenomena. First, given its poor cycling efficiency, the cathode rapidly forms a mixture of Li$_x$NMC at different lithiated states, having very high resistivity variations (i.e., 10$^{-7}$ Scm$^{-1}$ for $x$ = 0 and 10$^{-2}$ Scm$^{-1}$ for $x$ = 0.75, as reported by Amin et al. [25]). Second, the reactivity of the oxide cathode with the sulfide electrolyte leads to a resistive interface that accounts for the higher resistance, low coulombic efficiency, and severe capacity fading. This is not the case for TiS$_2$, for which better chemical compatibility between the cathode and electrolyte is confirmed by the lower interfacial resistance. Finally, we noticed that the Warburg impedance of the cathode and SE/cathode interface of the as-mounted I-SE cell was much higher than that of the Cl-SE one. This might indicate the formation of a passive interface due to a chemical reaction between I-SE and NMC as soon as they are in contact.

The electronic conductivity of the In-Li alloy was of the order of $10^{-6}$ Ohm cm and therefore the resistance Zw1 was mainly attributed to Li diffusion at the anode-electrolyte interface. For NMC, the values were around 1.1 kΩ increasing up to about 1.8 kΩ after 20 cycles, independently of the SE nature (Cl-SE or I-SE). Surprisingly, although the cathode chemistry was not expected to play a role at the anode/SE interface, the values of Zw1 appeared lower for $TiS_2$ (170 and 118 Ω, respectively). The main difference between cell impedances remained for the cathode counterpart.

Recently, solid-state batteries using metallic or complex hydrides as anodes or electrolytes have been reported [22]. Most of them use Li or In-Li alloy as anodes (except for López-Aranguren et al. [38] using a $MgH_2$-$TiH_2$ hydride). Therefore, the main difference between those InLi-anode batteries is linked to the electrolytes and the cathodes. The ones referring to oxides ($LiCoO_2$ or $Li_4Ti_5O_{12}$) [39,40] show rather poor reversible capacities (between 10 to 86 mAhg$^{-1}$) and need to be heated above the structural transition temperatures (60–120 °C) of the $LiBH_4$-like electrolytes to become significantly conductive. In the present work, thanks to the RT-conductive electrolytes developed by Dao et al. [23], the NMC-SE cells provided capacity around 75–100 mAhg$^{-1}$ at RT, even reaching 173 mAhg$^{-1}$ at 40 °C only. However, due to interfacial reaction between the cathode and the electrolyte, capacity fading occurred, showing the crucial point of compatibility between oxide cathodes and sulfide electrolytes. Indeed, Takahashi et al. [39] or Unemoto et al. [41] used $Li_3PO_4$-coated $LiCoO_2$ to prevent such interfacial reaction. Much better results were obtained with $TiS_2$. Though the use of high temperatures (80 to 130 °C) remains necessary for electrolytes based on borohydride or (car)boranes ($Li_2B_{12}H_{12}$, $LiCB_{11}H_{12}$) [41–44], lower working temperatures can be achieved for sulfide-based electrolytes ($Li(BH_4)_{0.75}I_{0.25}$ + $(Li_2S)_{0.75}(P_2S_5)_{0.25}$) [37,45]. Good capacities at RT are reported by these authors, but only for a limited number of cycles (less than ten). Moreover, nominal capacity of $TiS_2$ is not fully achieved and reversible capacities remain below 228 mAhg$^{-1}$. In the present work, using our new $(BH_4)^-$ substituted argyrodites, we reach 240 mAhg$^{-1}$ at RT (i.e., the nominal capacity of $TiS_2$) despite the need for two to three activation cycles related to the solid-electrolyte interface formation. Then, the battery provided full capacity for more than 35 cycles without fading, making a breakthrough for this type of RT titanium disulfide-based solid-state batteries.

## 5. Conclusions

In summary, using a new class of $(BH_4)^-$ substituted argyrodites $Li_6PS_5Z_{0.83}(BH_4)_{0.17}$, (Z = Cl, I), room temperature solid-state batteries have been set up by combining NMC or $TiS_2$ cathodes with In-Li anodes. The use of oxide cathode allowed developing batteries with high voltage and good nominal capacity, but the cell suffered from interfacial reaction at the cathode side, leading to a strong resistance increase and fading of the reversible capacity over cycling. This might have been overcome by coating the active material with a protective layer (e.g., $LiNbO_3$) [30]. Using a $TiS_2$ cathode, though presenting lower voltage, a higher capacity was achieved (240 mAhg$^{-1}$) and full reversibility of the cathode was observed and preserved upon cycling with CE above 97%. This was attributed to the good compatibility between $TiS_2$ and argyrodites, preventing the formation of an insulating layer at the cathode side. Energy density can still be improved by increasing the cell voltage using a low potential anode (made of metal hydrides for example) instead of the In-Li one. Improvements are still needed regarding lithium conductivity at RT that remains a limiting factor for high *C*-rates.

writing—review and editing, A.H.D., P.L.-A., J.Z., F.C., and M.L.; supervision, F.C. and M.L.; funding acquisition, F.C. and M.L. All authors have read and agreed to the published version of the manuscript.

**Funding:** This research was funded from the European Union's Seventh Framework Program for research technological development and demonstration under grant agreement no. 607040 through the Marie Curie ITN ECOSTORE project.

**Acknowledgments:** The Ministerio de Ciencia, Innovación y Universidades of Spain is greatly acknowledged for the Juan de la Cierva grant under the reference IJCI-2017-32310. Authors would like to thank SAFT for technical support.

**Conflicts of Interest:** The authors declare no conflict of interest.

**References**

1. Dudney, N., J.; West, W.C.; Nanda, J. *Handbook of Solid State Batteries, 2nd Edition.*; 2nd Edition.; World Scientific : Singapore, 2015; ISBN 978-981-4651-89-9.
2. Gao, Z.; Sun, H.; Fu, L.; Ye, F.; Zhang, Y.; Luo, W.; Huang, Y. Promises, Challenges, and Recent Progress of Inorganic Solid-State Electrolytes for All-Solid-State Lithium Batteries. *Advanced Materials* **2018**, *30*, 1705702, doi:10.1002/adma.201705702.
3. Kim, J.G.; Son, B.; Mukherjee, S.; Schuppert, N.; Bates, A.; Kwon, O.; Choi, M.J.; Chung, H.Y.; Park, S. A review of lithium and non-lithium based solid state batteries. *Journal of Power Sources* **2015**, *282*, 299–322, doi:10.1016/j.jpowsour.2015.02.054.
4. Takada, K. Progress and Prospective of Solid-State Lithium Batteries. *Acta Materialia* **2013**, *61*, 759–770, doi:10.1016/j.actamat.2012.10.034.
5. Lau, J.; DeBlock, R.H.; Butts, D.M.; Ashby, D.S.; Choi, C.S.; Dunn, B.S. Sulfide Solid Electrolytes for Lithium Battery Applications. *Advanced Energy Materials* **2018**, *8*, 1800933, doi:10.1002/aenm.201800933.
6. Kato, Y.; Hori, S.; Saito, T.; Suzuki, K.; Hirayama, M.; Mitsui, A.; Yonemura, M.; Iba, H.; Kanno, R. High-power all-solid-state batteries using sulfide superionic conductors. *Nature Energy* **2016**, *1*, 16030, doi:10.1038/nenergy.2016.30.
7. Sulfide Solid Electrolyte with Favorable Mechanical Property for All-Solid-State Lithium Battery | Scientific Reports Available online: https://www.nature.com/articles/srep02261 (accessed on Apr 23, 2020).
8. Randau, S.; Weber, D.A.; Kötz, O.; Koerver, R.; Braun, P.; Weber, A.; Ivers-Tiffée, E.; Adermann, T.; Kulisch, J.; Zeier, W.G.; et al. Benchmarking the performance of all-solid-state lithium batteries. *Nature Energy* **2020**, *5*, 259–270, doi:10.1038/s41560-020-0565-1.
9. Janek, J.; Zeier, W.G. A solid future for battery development. *Nature Energy* **2016**, *1*, 16141, doi:10.1038/nenergy.2016.141.
10. Kitaura, H.; Hayashi, A.; Tadanaga, K.; Tatsumisago, M. Electrochemical performance of all-solid-state lithium secondary batteries with Li–Ni–Co–Mn oxide positive electrodes. *Electrochimica Acta* **2010**, *55*, 8821–8828, doi:10.1016/j.electacta.2010.07.066.
11. Machida, N.; Kashiwagi, J.; Naito, M.; Shigematsu, T. Electrochemical properties of all-solid-state batteries with ZrO2-coated LiNi$_{1/3}$Mn$_{1/3}$Co$_{1/3}$O$_2$ as cathode materials. *Solid State Ionics* **2012**, *225*, 354–358, doi:10.1016/j.ssi.2011.11.026.
12. Okada, K.; Machida, N.; Naito, M.; Shigematsu, T.; Ito, S.; Fujiki, S.; Nakano, M.; Aihara, Y. Preparation and electrochemical properties of LiAlO2-coated Li(Ni$_{1/3}$Mn$_{1/3}$Co$_{1/3}$)O$_2$ for all-solid-state batteries. *Solid State Ionics* **2014**, *255*, 120–127, doi:10.1016/j.ssi.2013.12.019.

13. Meng, H.; Huang, B.; Yin, J.; Yao, X.; Xu, X. Synthesis and electrochemical properties of LiNi$_{1/3}$Co$_{1/3}$Mn1/3O$_2$ cathodes in lithium-ion and all-solid-state lithium batteries. *Ionics* **2015**, *21*, 43–49, doi:10.1007/s11581-014-1152-x.

14. Deiseroth, H.-J.; Kong, S.-T.; Eckert, H.; Vannahme, J.; Reiner, C.; Zaiß, T.; Schlosser, M. Li6PS5X: A Class of Crystalline Li-Rich Solids With an Unusually High Li+ Mobility. *Angewandte Chemie International Edition* **2008**, *47*, 755–758, doi:10.1002/anie.200703900.

15. Jung, W.D.; Kim, J.-S.; Choi, S.; Kim, S.; Jeon, M.; Jung, H.-G.; Chung, K.Y.; Lee, J.-H.; Kim, B.-K.; Lee, J.-H.; et al. Superionic Halogen-Rich Li-Argyrodites Using In Situ Nanocrystal Nucleation and Rapid Crystal Growth. *Nano Lett.* **2020**, *20*, 2303–2309, doi:10.1021/acs.nanolett.9b04597.

16. Koerver, R.; Aygün, I.; Leichtweiß, T.; Dietrich, C.; Zhang, W.; Binder, J.O.; Hartmann, P.; Zeier, W.G.; Janek, J. Capacity Fade in Solid-State Batteries: Interphase Formation and Chemomechanical Processes in Nickel-Rich Layered Oxide Cathodes and Lithium Thiophosphate Solid Electrolytes. *Chem. Mater.* **2017**, *29*, 5574–5582, doi:10.1021/acs.chemmater.7b00931.

17. Li, W.J.; Hirayama, M.; Suzuki, K.; Kanno, R. Fabrication and All Solid-State Battery Performance of TiS$_2$/Li$_{10}$GeP$_2$S$_{12}$ Composite Electrodes. *Materials Transactions* **2016**, *57*, 549–552, doi:10.2320/matertrans.Y-M2016804.

18. Shin, B.R.; Nam, Y.J.; Oh, D.Y.; Kim, D.H.; Kim, J.W.; Jung, Y.S. Comparative Study of TiS2/Li-In All-Solid-State Lithium Batteries Using Glass-Ceramic Li3PS4 and Li10GeP2S12 Solid Electrolytes. *Electrochimica Acta* **2014**, *146*, 395–402, doi:10.1016/j.electacta.2014.08.139.

19. Auvergniot, J.; Cassel, A.; Ledeuil, J.-B.; Viallet, V.; Seznec, V.; Dedryvère, R. Interface Stability of Argyrodite Li$_6$PS$_5$Cl toward LiCoO$_2$, LiNi$_{1/3}$Co$_{1/3}$Mn$_{1/3}$O$_2$, and LiMn$_2$O$_4$ in Bulk All-Solid-State Batteries. *Chem. Mater.* **2017**, *29*, 3883–3890, doi:10.1021/acs.chemmater.6b04990.

20. de Jongh, P.E.; Blanchard, D.; Matsuo, M.; Udovic, T.J.; Orimo, S. Complex hydrides as room-temperature solid electrolytes for rechargeable batteries. *Applied Physics A* **2016**, *122*, 251, doi:10.1007/s00339-016-9807-2.

21. Matsuo, M.; Orimo, S. Lithium Fast-Ionic Conduction in Complex Hydrides: Review and Prospects. *Advanced Energy Materials* **2011**, *1*, 161–172, doi:10.1002/aenm.201000012.

22. Latroche, M.; Blanchard, D.; Cuevas, F.; El Kharbachi, A.; Hauback, B.C.; Jensen, T.R.; de Jongh, P.E.; Kim, S.; Nazer, N.S.; Ngene, P.; et al. Full-cell hydride-based solid-state Li batteries for energy storage. *International Journal of Hydrogen Energy* **2019**, *44*, 7875–7887, doi:10.1016/j.ijhydene.2018.12.200.

23. Ha Dao, A.; López-Aranguren, P.; Černý, R.; Guiader, O.; Zhang, J.; Cuevas, F.; Latroche, M.; Jordy, C. Improvement of the ionic conductivity on new substituted borohydride argyrodites. *Solid State Ionics* **2019**, *339*, 114987, doi:10.1016/j.ssi.2019.05.022.

24. Whittingham, M.S. Lithium Batteries and Cathode Materials. *Chem. Rev.* **2004**, *104*, 4271–4302, doi:10.1021/cr020731c.

25. Amin, R.; Chiang, Y.-M. Characterization of Electronic and Ionic Transport in Li$_{1-x}$Ni$_{0.33}$Mn$_{0.33}$Co$_{0.33}$O$_2$ (NMC333) and Li$_{1-x}$Ni$_{0.50}$Mn$_{0.20}$Co$_{0.30}$O$_2$ (NMC523) as a Function of Li Content. *Journal of The Electrochemical Society* **2016**, *163*, A1512–A1517.

26. Logothetis, E.M.; Kaiser, W.J.; Kukkonen, C.A.; Faile, S.P.; Colella, R.; Gambold, J. Hall coefficient and reflectivity evidence that TiS2 is a semiconductor. *Journal of Physics C: Solid State Physics* **1979**, *12*, L521–L526, doi:10.1088/0022-3719/12/13/007.

27. Julien, C.; Nazri, G.-A. *Solid State Batteries: Materials Design and Optimization*; The Springer International Series in Engineering and Computer Science; Springer US: New York, NY, USA, 1994; Vol. 271; ISBN 978-0-7923-9460-0.

28. Kaiser, N.; Spannenberger, S.; Schmitt, M.; Cronau, M.; Kato, Y.; Roling, B. Ion transport limitations in all-solid-state lithium battery electrodes containing a sulfide-based electrolyte. *Journal of Power Sources* **2018**, *396*, 175–181, doi:10.1016/j.jpowsour.2018.05.095.

29. Santhosha, A.L.; Medenbach, L.; Buchheim, J.R.; Adelhelm, P. The Indium−Lithium Electrode in Solid-State Lithium-Ion Batteries: Phase Formation, Redox Potentials, and Interface Stability. *Batteries & Supercaps* **2019**, *2*, 524–529, doi:10.1002/batt.201800149.

30. Ohta, N.; Takada, K.; Sakaguchi, I.; Zhang, L.; Ma, R.; Fukuda, K.; Osada, M.; Sasaki, T. LiNbO$_3$-coated LiCoO$_2$ as cathode material for all solid-state lithium secondary batteries. *Electrochemistry Communications* **2007**, *9*, 1486–1490, doi:10.1016/j.elecom.2007.02.008.

31. Ohta, N.; Takada, K.; Zhang, L.; Ma, R.; Osada, M.; Sasaki, T. Enhancement of the High-Rate Capability of Solid-State Lithium Batteries by Nanoscale Interfacial Modification. *Advanced Materials* **2006**, *18*, 2226–2229, doi:10.1002/adma.200502604.

32. Haruyama, J.; Sodeyama, K.; Tateyama, Y. Cation Mixing Properties toward Co Diffusion at the LiCoO$_2$ Cathode/Sulfide Electrolyte Interface in a Solid-State Battery. *ACS Appl. Mater. Interfaces* **2017**, *9*, 286–292, doi:10.1021/acsami.6b08435.

33. Ohtomo, T.; Hayashi, A.; Tatsumisago, M.; Tsuchida, Y.; Hama, S.; Kawamoto, K. All-solid-state lithium secondary batteries using the 75Li$_2$S·25P$_2$S$_5$ glass and the 70Li$_2$S·30P$_2$S$_5$ glass–ceramic as solid electrolytes. *Journal of Power Sources* **2013**, *233*, 231–235, doi:10.1016/j.jpowsour.2013.01.090.

34. Sakuda, A.; Hayashi, A.; Tatsumisago, M. Interfacial Observation between LiCoO$_2$ Electrode and Li$_2$S−P$_2$S$_5$ Solid Electrolytes of All-Solid-State Lithium Secondary Batteries Using Transmission Electron Microscopy. *Chem. Mater.* **2010**, *22*, 949–956, doi:10.1021/cm901819c.

35. Woo, J.H.; Trevey, J.E.; Cavanagh, A.S.; Choi, Y.S.; Kim, S.C.; George, S.M.; Oh, K.H.; Lee, S.-H. Nanoscale Interface Modification of LiCoO$_2$ by Al$_2$O$_3$ Atomic Layer Deposition for Solid-State Li Batteries. *Journal of The Electrochemical Society* **2012**, *159*, A1120–A1124.

36. Sakuda, A.; Yamauchi, A.; Yubuchi, S.; Kitamura, N.; Idemoto, Y.; Hayashi, A.; Tatsumisago, M. Mechanochemically Prepared Li2S–P2S5–LiBH4 Solid Electrolytes with an Argyrodite Structure. *ACS Omega* **2018**, *3*, 5453–5458, doi:10.1021/acsomega.8b00377.

37. Unemoto, A.; Wu, H.; Udovic, T.J.; Matsuo, M.; Ikeshoji, T.; Orimo, S. Fast lithium-ionic conduction in a new complex hydride–sulphide crystalline phase. *Chem. Commun.* **2016**, *52*, 564–566, doi:10.1039/C5CC07793A.

38. López-Aranguren, P.; Berti, N.; Dao, H.A.; Zhang, J.; Cuevas, F.; Latroche, M.; Jordy, C. An all-solid-state metal hydride – Sulfur lithium-ion battery. *J. Power Sources* **2017**, *357*, 56–60, doi:10.1016/j.jpowsour.2017.04.088.

39. Takahashi, K.; Hattori, K.; Yamazaki, T.; Takada, K.; Matsuo, M.; Orimo, S.; Maekawa, H.; Takamura, H. All-solid-state lithium battery with LiBH$_4$ solid electrolyte. *Journal of Power Sources* **2013**, *226*, 61–64, doi:10.1016/j.jpowsour.2012.10.079.

40. Sveinbjörnsson, D.; Christiansen, A.S.; Viskinde, R.; Norby, P.; Vegge, T. The LiBH$_4$-LiI Solid Solution as an Electrolyte in an All-Solid-State Battery. *Journal of The Electrochemical Society* **2014**, *161*, A1432–A1439.

41. Unemoto, A.; Nogami, G.; Tazawa, M.; Taniguchi, M.; Orimo, S. Development of 4V-Class Bulk-Type All-Solid-State Lithium Rechargeable Batteries by a Combined Use of Complex Hydride and Sulfide Electrolytes for Room Temperature Operation. *Materials Transactions* **2017**, *58*, 1063–1068, doi:10.2320/matertrans.M2017022.

42. Unemoto, A.; Ikeshoji, T.; Yasaku, S.; Matsuo, M.; Stavila, V.; Udovic, T.J.; Orimo, S. Stable Interface Formation between TiS$_2$ and LiBH$_4$ in Bulk-Type All-Solid-State Lithium Batteries. *Chem. Mater.* **2015**, *27*, 5407–5416, doi:10.1021/acs.chemmater.5b02110.

43. Kim, S.; Toyama, N.; Oguchi, H.; Sato, T.; Takagi, S.; Ikeshoji, T.; Orimo, S. Fast Lithium-Ion Conduction in Atom-Deficient *closo*-Type Complex Hydride Solid Electrolytes. *Chem. Mater.* **2018**, *30*, 386–391, doi:10.1021/acs.chemmater.7b03986.

44. Tang, W.S.; Matsuo, M.; Wu, H.; Stavila, V.; Zhou, W.; Talin, A.A.; Soloninin, A.V.; Skoryunov, R.V.; Babanova, O.A.; Skripov, A.V.; et al. Liquid-Like Ionic Conduction in Solid Lithium and Sodium Monocarba-*closo*-Decaborates Near or at Room Temperature. *Advanced Energy Materials* **2016**, *6*, 1502237, doi:10.1002/aenm.201502237.

45. El Kharbachi, A.; Hu, Y.; Sørby, M.H.; Mæhlen, J.P.; Vullum, P.E.; Fjellvåg, H.; Hauback, B.C. Reversibility of metal-hydride anodes in all-solid-state lithium secondary battery operating at room temperature. *Solid State Ionics* **2018**, *317*, 263–267, doi:10.1016/j.ssi.2018.01.037.

# Solid-state Li-ion batteries operating at room temperature using new borohydride argyrodite electrolytes.

# Supplementary Information (SI)

**Anh Ha Dao [1], Pedro López-Aranguren [1,2], Junxian Zhang [1], Fermín Cuevas [1] and Michel Latroche [1]**

[1] department/school/faculty/campus, Univ. Paris Est Creteil, CNRS, ICMPE, UMR7182, 7182, 2 rue Henri Dunant, F-94320 Thiais, France; daohaanh1988@gmail.com (A.H.D.); plopez@cicenergigune.com (P.L.-A.); junxian@icmpe.cnrs.fr (J.Z.); latroche@icmpe.cnrs.fr (M.L.)

[2] Center for Cooperative Research on Alternative Energies (CIC energiGUNE), Basque Research and Technology Alliance (BRTA), Parque Tecnológico de Álava, Albert Einstein, 48, 01510 Vitoria-Gasteiz, Spain

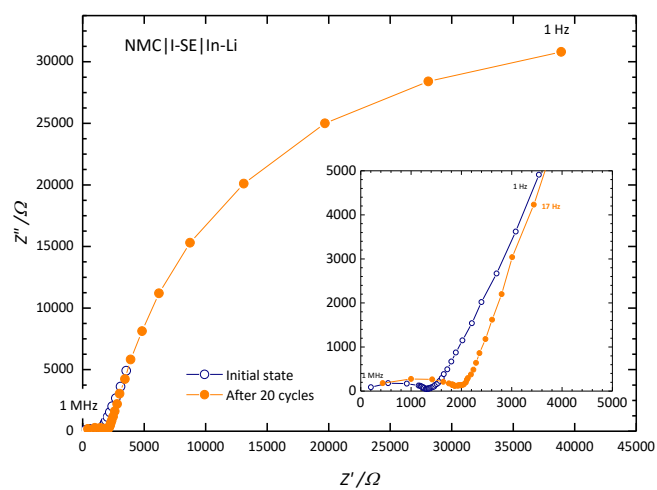

Figure SI1 : Impedance profiles of the NMC|I-SE|In-Li cell obtained before (open purple circle) and after (full orange circle) 20 cycles at 25ºC.